\documentclass[a4paper]{jpconf}
\usepackage{graphicx}
\usepackage{graphics} 
\usepackage{xcolor} 
\usepackage{amsmath} 
\usepackage{amssymb} 
\usepackage[normalem]{ulem} 
\usepackage{cite} 
\usepackage{lineno} 

\begin{document}
\title{Calculated energy loss of swift light ions in platinum and gold: importance of the target electronic excitation spectrum}

\author{Isabel Abril$^1$, Pablo de Vera$^2$ and Rafael Garcia-Molina$^2$}

\address{$^1$ Departament de Física Aplicada, Universitat d’Alacant, 03080 Alacant, Spain}

\address{$^2$ Departamento de Física - Centro de Investigación en \'Optica y Nanof\'isica, Universidad de Murcia, 30100 Murcia, Spain}

\ead{rgm@um.es}

\begin{abstract}
Understanding and predicting the energy loss of swift ions in metals is important for many applications of charged particle beams, such as analysis and modification of materials, and recently for modelling metal nanoparticle radiosensitisation in ion beam cancer therapy. We have calculated the stopping power of the transition metals Pt and Au for protons and alpha particles in a wide energy range, using the dielectric formalism, which realistically accounts for the excitation spectrum of each metal through the Mermin Energy Loss Function - Generalised Oscillator Strength methodology. 
For each combination of projectile, energy and target, we have considered: (i) the equilibrium charge state of the projectile through the target, (ii) the energy-loss due to electron capture and loss processes, and (iii) the energy loss resulting from the polarisation of the projectile's electronic cloud due to the self-induced electric field. 
Our calculated stopping powers show a fairly good agreement with the available experimental data for platinum and gold, particularly the most recent ones around the stopping power maximum, which validates the methodology we have used to be further extended to other transition metals. 
For the materials studied (platinum and gold), two commonly used and different sources of the experimental excitation spectrum yield comparable calculated stopping powers and mean excitation energies, the latter being closer to the most recent data provided in a recent ICRU Report than to previous compilations.
Despite the small differences in the sources of excitation spectra of these metals, they lead to practically the same stopping power results as far as they reproduce the main excitation features of the material and fulfil physically motivated sum rules.

\end{abstract}

\section{Introduction}
The interaction of charged particles with targets of different nature (metals, insulators, biomaterials…) is important in order to know in detail how the projectile transfers energy to the target components and, therefore, improve the techniques to modify or characterise materials for a wide range of applications \cite{Nastasi1996,Sigmund2006}.

Besides the previously mentioned interest of better describing the interaction of radiation with matter (of any type) for materials analysis and modification, transition metals are becoming widely used in nanoparticles to improve cancer treatment through radiotherapy \cite{Porcel2010,Lacombe2017,Kuncic2018,Kempson2021}.  
One of the particularities of transition metals is their capability of generating a large number of low-energy electrons when irradiated \cite{Verkhovtsev2015e}.
These electrons are considered to produce important (bio)damage at the cellular level \cite{Zheng2008-RadRes}.
Therefore, it is expected that addition of nanoparticles (of gold, platinum or other transition metals) targeting the tumour region will improve the efficiency of cancer radiotherapy treatments. 


In this work, we have used the dielectric formalism to calculate the average electronic energy loss per unit path length (stopping power) for proton and alpha particle beams in the condensed-phase metals platinum and gold commonly used for radioenhancing nanoparticles. For that purpose, two common experimental sets for the optical electronic excitation spectrum of the metals (one based on a compilation of optical data \cite{Palik1999} and other on reflection electron energy loss spectroscopy, REELS \cite{Werner2009}) have been used as input for the calculations. Both sets of data yield fairly similar results, and in good agreement with the most recent experimental determinations of the energy loss quantities for light ion beams.

\section{Dielectric formalism}
The generation of secondary electrons depends on the energy deposited in the nanoparticles by the incident projectiles (protons, alpha particles, carbon ions...). 
The probability that a charged projectile (with energy $T$, mass $M$, atomic number $Z$, and charge state $q$) transfers momentum $\hbar k$ and energy $\hbar \omega$ to a material can be obtained from the dielectric formalism \cite{Lindhard1954,Ritchie1959}, where the characteristics of the projectile and those of the target appear separately in the doubly differential cross section:
\begin{equation}
	\frac{\mathrm{d}^2 \Lambda_q (T)}{\mathrm{d} \omega \mathrm{d} k} = \frac{e^2}{\hbar^2 \pi} \frac{M [Z - \rho_q(k)]^2}{T} \frac{1}{k} \mathrm{Im} \left[ \frac{-1}{\epsilon(k, \omega)} \right] ,
\label{eq:DDCS}
\end{equation}
where $\rho_q(k)$ is the Fourier transform of the projectile's electron density, and $\mathrm{Im} \left[ -1/\epsilon(k, \omega) \right]$ is the energy loss function (ELF) of the target, which contains information about its electronic excitation spectrum; $\epsilon(k, \omega)$ is the complex dielectric function of the material.

From eq. \eqref{eq:DDCS}, all other useful energy loss quantities can be obtained, such as the inverse mean free path: 
\begin{equation}
	\Lambda_q (T) = \int_{\omega_-}^{\omega_+} \mathrm{d} \omega \int_{k_-}^{k_+} \mathrm{d} k \frac{\mathrm{d}^2 \Lambda_q (T)}{\mathrm{d} \omega \mathrm{d} k}, 
	\label{eq:Lambda_q}
\end{equation}
the stopping power:
\begin{equation}
	S_q (T) = \int_{\omega_-}^{\omega_+} \mathrm{d} \omega \, \hbar \omega \int_{k_-}^{k_+} \mathrm{d} k \frac{\mathrm{d}^2 \Lambda_q (T)}{\mathrm{d} \omega \mathrm{d} k}, 
\label{eq:S_q}
\end{equation}
or the energy-loss straggling: 
\begin{equation}
	\Omega^2_q (T) = \int_{\omega_-}^{\omega_+} \mathrm{d} \omega \, (\hbar \omega)^2  \int_{k_-}^{k_+} \mathrm{d} k \frac{\mathrm{d}^2 \Lambda_q (T)}{\mathrm{d} \omega \mathrm{d} k}.
\label{eq:Omega2_q}
\end{equation}
The limits in the integrations correspond to the minimum and maximum energy and momentum transfers, which can be deduced from the allowed energy transitions of the target electrons and conservation laws.




To evaluate these integrals, the ELF of the target has to be known as a function of $\hbar k$ and $\hbar \omega$.  
For most condensed media, the ELF is not available for all $\hbar k$ and $\hbar \omega$. However, there is a considerable amount of optical data, in the limit $k=0$, which are extended to $k>0$ through suitable dispersion relations \cite{GarciaMolina2012SpringerScience}.

The doubly differential cross section, eq. \eqref{eq:DDCS}, and hence the energy loss function is also a basic input for Monte Carlo codes simulating radiation transport in condensed-phase biological matter. 
A large number of Monte Carlo studies of the track structure of electrons (as well as positrons or heavy ions) in water have been conducted. However, one current difficulty with performing these studies for metals is the need of reliable, realistic cross sections for the condensed phase.

\section{Energy loss functions of platinum and of gold}
In order to evaluate the integrals appearing in eqs. (\ref{eq:Lambda_q}, \ref{eq:S_q}, \ref{eq:Omega2_q}), it is convenient to have an analytical expression for the ELF in the whole $k$-$\omega$ plane (the so-called Bethe surface). This is obtained by using the MELF-GOS methodology \cite{HerediaAvalos2005PRA,GarciaMolina2011}, through a sum of Mermin-type \cite{Mermin1970} ELF for the excitation of the target's outer-shell electrons ($\hbar \omega \lesssim 100$ eV), Generalised Oscillator Strengths for the inner-shell electrons ($\hbar \omega \gtrsim 1000$ eV), and semi-core levels in between described by Mermin-type ELF. Proceeding in this manner, the optical (i.e., at $k=0$) ELF can be easily obtained from experiments, which is straightforwardly extended to finite momentum transfers (i.e. $k \neq 0$) thanks to the analytical properties of the Mermin dielectric function \cite{Mermin1970} and the Generalised Oscillator Strengths.
  
\begin{figure}[h]
	\includegraphics[width=18.5pc]{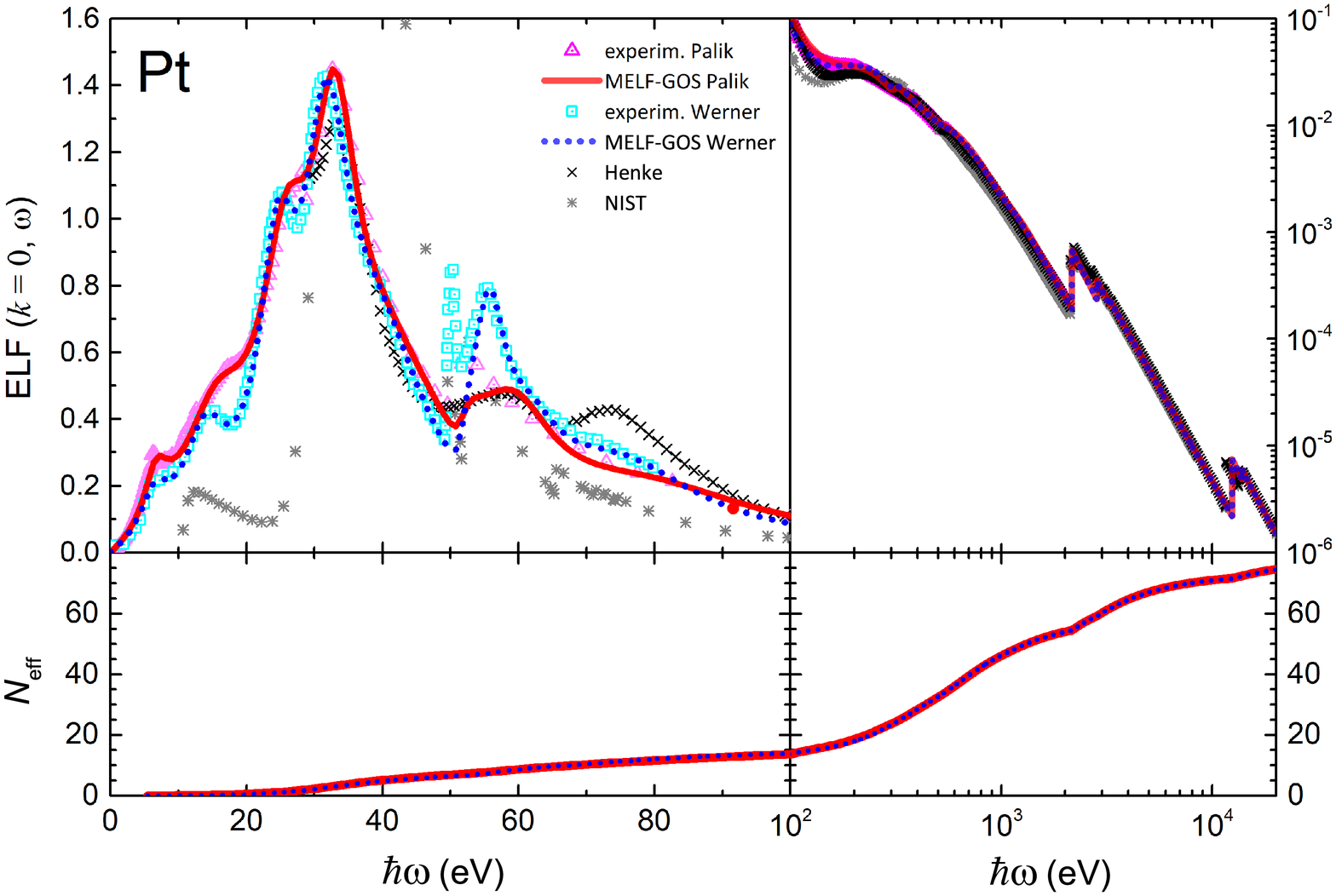} \quad
	\includegraphics[width=18.5pc]{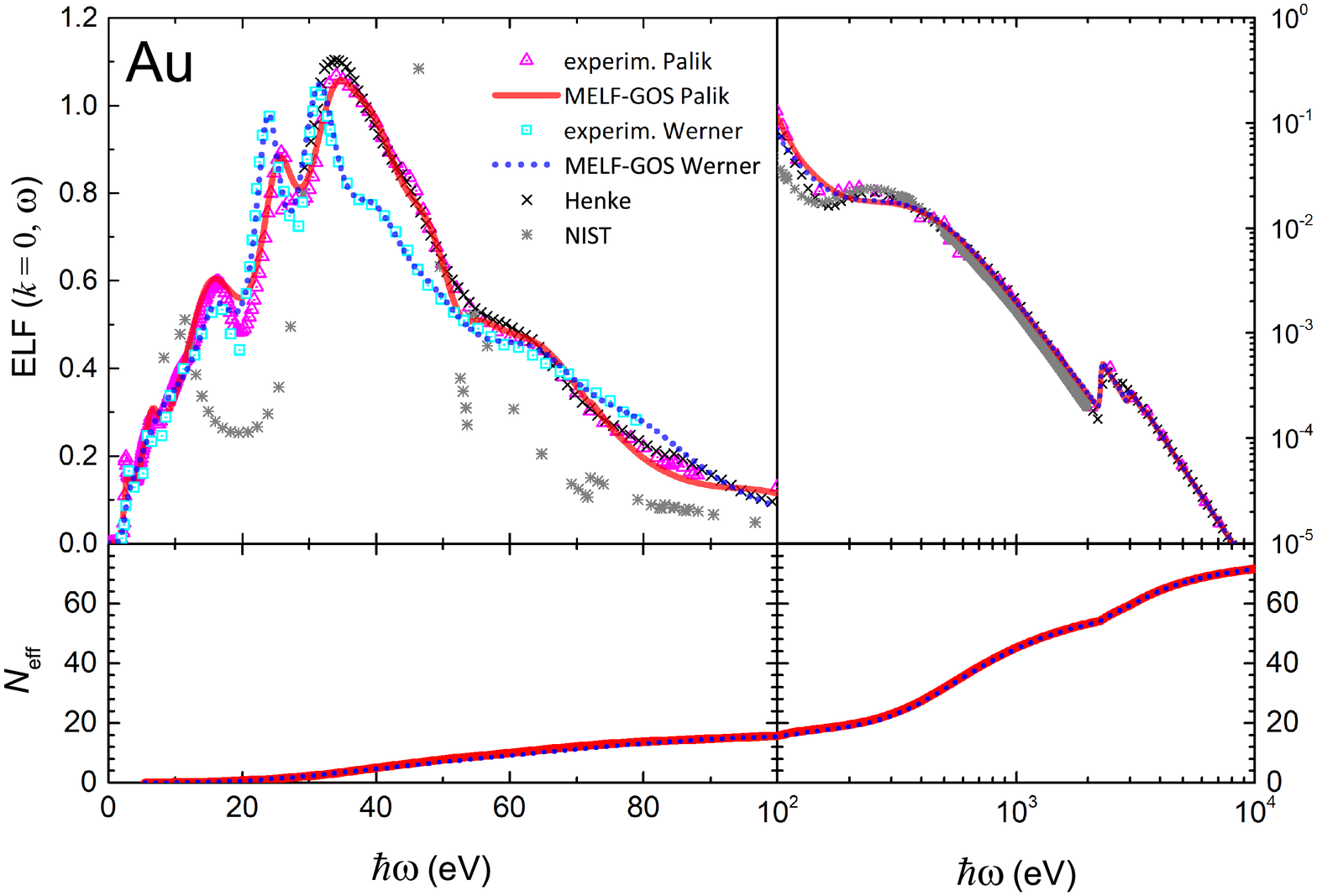}
	\caption{The upper panels show the optical (i.e. $k=0$) energy loss functions (ELF) of platinum and gold, as a function of the excitation energy $\hbar \omega$. Symbols and lines correspond, respectively, to experimental data and MELF-GOS fittings (triangles and continuous line from Ref. \cite{Palik1999}, circles and dotted line from Ref. \cite{Werner2009}). At high energies the results from Refs. \cite{Henke1993,NIST-Chantler2003} are also depicted. The lower panels show the number of effective electrons of each material, as a function of the excitation energy. \label{fig:ELF-Neff}}
\end{figure}

Symbols in figure \ref{fig:ELF-Neff} represent the optical ELF of platinum and gold, derived from experimental \cite{Palik1999,Werner2009} and calculated \cite{Henke1993,NIST-Chantler2003} values of the  the most external and most internal, respectively, excitations of each material.
The continuous and dotted curves in the upper panels of figure \ref{fig:ELF-Neff} represent the analytical optical ELF obtained from the MELF-GOS method and fitted, respectively, to optical (triangle symbols \cite{Palik1999}) and to REELS (square symbols \cite {Werner2009}) experimental data. At high energy transfers, a comparison with results obtained from X-ray atomic scattering factors \cite{Henke1993, NIST-Chantler2003} is also presented.

The analytical ELF should not only reproduce (reasonably) well the structure of the experimental ELF, but it also must fulfil certain physical rules, such as the $f$-sum rule \cite{Shiles1980}, which gives the effective number of electrons $N_\mathrm{eff}$ per target atom that are involved in excitations up to an energy $\hbar \omega$:
\begin{equation}
	N_{\rm eff}(\omega) = \frac{m}{2\pi^2 e^2 \mathcal{N}} \int_0^\omega \mathrm{d} \omega' \, \omega' \, \textrm{Im}\left[ \frac{-1}{\epsilon(0, \omega')} \right] \mbox{ , }
	\label{eq:f-sum_ELF}
\end{equation}
where $m$ is the electron mass and $\mathcal{N}$ is the atomic density of the target.
The lower panels of figure \ref{fig:ELF-Neff} show how $N_\mathrm{eff}$ evolves as the excitation energy $\hbar \omega$ increases, tending to the total number of electrons of platinum ($Z=78$) and gold ($Z=79$) as $\hbar \omega \rightarrow \infty$. 

An important property of each metal, useful in the calculation of the energy loss of charged particles through matter, is the mean excitation energy $I$, which is the main parameter determining the stopping power at high ion energies through the Bethe equation \cite{Fano1963}. The mean excitation energy is obtained from the ELF by means of the following expression \cite{Fano1963,Smith2006}:
\begin{equation}
	\ln I = \frac{\int_0^\infty \mathrm{d}  \omega \,  \omega \ln (\hbar \omega) \mathrm{Im} [-1/\epsilon(k=0, \omega)]}{\int_0^\infty \mathrm{d}  \omega \,  \omega \, \mathrm{Im} [-1/\epsilon(k=0, \omega)]} .
	\label{eq:I}
\end{equation}

Table \ref{tab:I} contains the mean excitation energies of platinum and gold obtained in the present work when using the energy loss function derived through the MELF-GOS method from the two sets of experimental data \cite{Palik1999, Werner2009}. These values of $I$ are rather similar for each element, independently of the origin of the experimental ELF (as far as it provides a realistic description of the target electronic excitation spectrum). The values of $I$ obtained in this work are much closer to the ones provided in a recent report \cite{ICRU-Report73} than to the values (790 eV for both metals) appearing in the Particle Data Group webpage \cite{ParticleDataGroup}. 

\begin{table}[h]
	\caption{Mean excitation energy $I$ of platinum and gold, as obtained in this work from two different sets of experimental data (optical measurementes \cite{Palik1999}, REELS \cite{Werner2009}), and in a recent publication \cite{ICRU-Report73}. \label{tab:I}
	} 
	\begin{center}
		\lineup
		\begin{tabular}{cccc}
			\br                             
		     $I$ (eV)	   & This work (ELF from Ref. \cite{Palik1999})   &  This work (ELF from Ref. \cite{Werner2009}) &  Ref. \cite{ICRU-Report73} \cr 
			\mr 
			Pt   & $738.6$ & $739.0$ & $751.6$ \cr       
			Au   & $761.2$ & $775.9$ & $741.9$ \cr
			\br
		\end{tabular}
	\end{center}
\end{table}

\section{Stopping power of platinum and gold for protons and alpha particles}
When a swift projectile moves through a material, it experiences an energy loss per unit path length, or retarding force, which is dubbed stopping power $S(T)$. This quantity is calculated taking into account that the projectile can change its charge state as it travels through a material by capturing and losing electrons. The processes of electron capture and loss also contribute to the energy loss. Finally, the effects due to the deformation of the projectile's electronic cloud due to the electric field induced in the target is also included in the calculations. All these terms add to give the stopping power $S(T)$:
\begin{equation}
	S(T) = \sum_q \phi_q S_q(T) + S_{\textsc{c\&l}} (T) ,
\end{equation}
where $\phi_q (T)$ is the probability of the projectile having a charge fraction $q$ \cite{Schiwietz2001}. The stopping power $S_q(T)$ corresponding to each charge state $q$ is modified to take into account the polarisation of the projectile \cite{HerediaAvalos2002Polarization}, and $S_{\textsc{c\&l}} (T)$ is the contribution from the capture and loss of electrons \cite{HerediaAvalos2002Polarization, Denton2008b}.



\begin{figure}[h]
	\includegraphics[height=25pc]{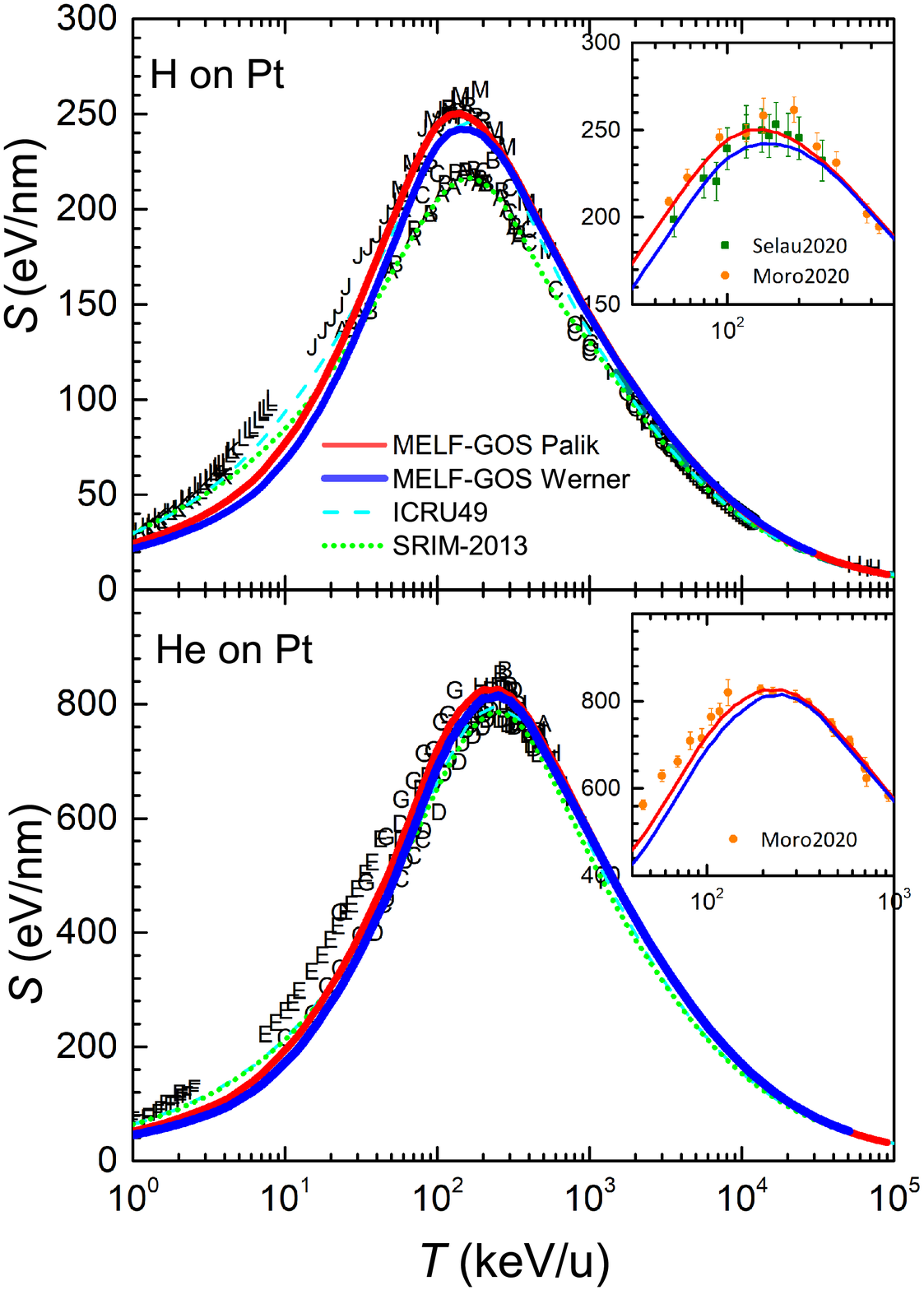} \quad
	\includegraphics[height=25pc]{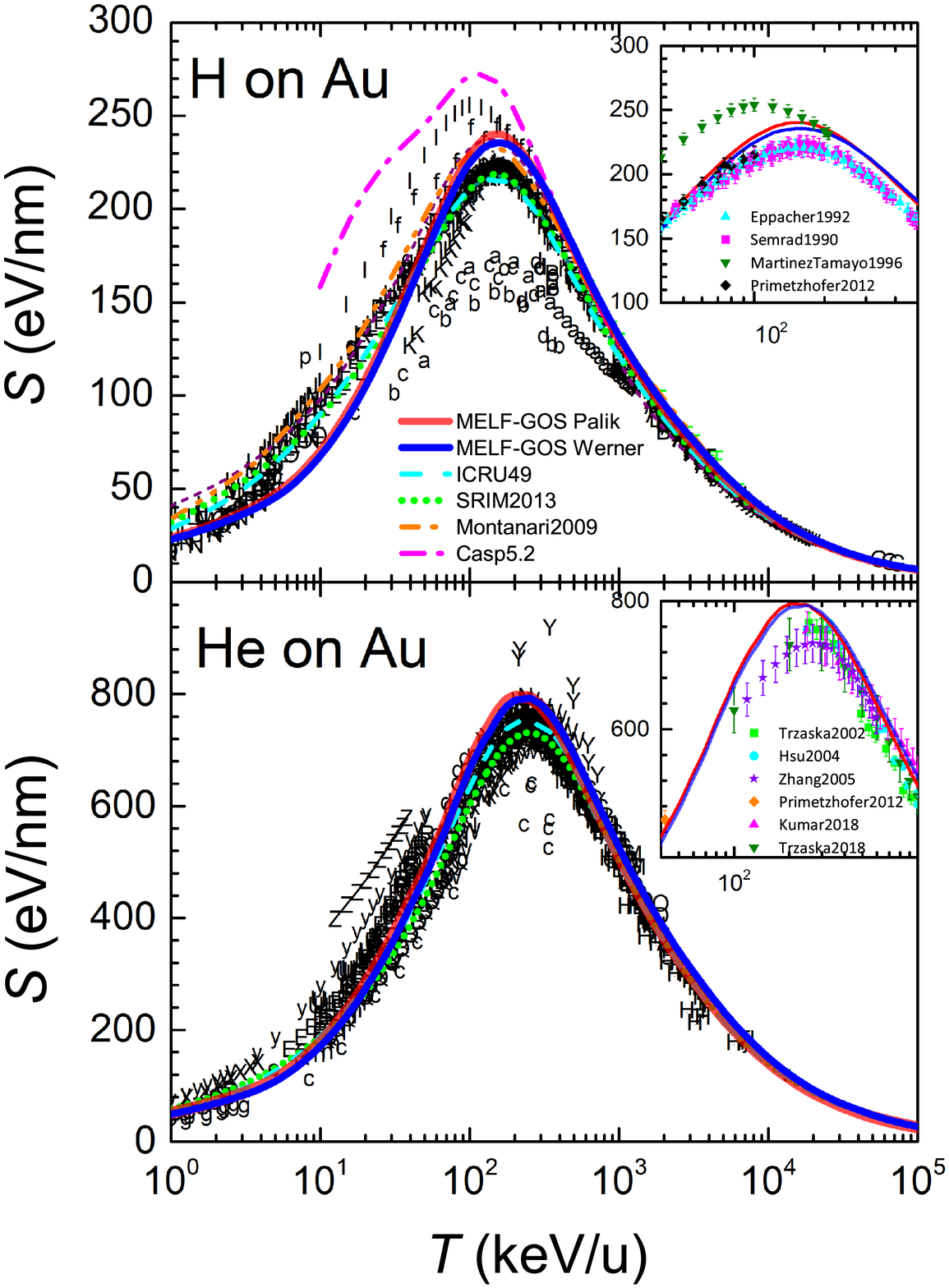}

	\caption{Stopping power of platinum (left panels) and gold (right panels) for protons (upper panels) and alpha particles (lower panels). Continuous curves are our results (red and blue: using the ELF derived from Refs. \cite{Palik1999} and \cite{Werner2009}, respectively). Discontinuous curves represent other results \cite{ICRU49,Schiwietz2012,SRIM2013,Montanari2013}. Experimental data are represented by letters \cite{PaulDatabase} and symbols \cite{Semrad1990,Eppacher1992,MartinezTamayo1996,Trzaska2002,Hsu2004,Zhang-Weber2005,Primetzhofer2012,Kumar2018,Trzaska2018,Moro2020,Selau2020}.  
	\label{fig:Sp-Pt-Au}}
\end{figure}

Proceeding in this manner, we have calculated the stopping power of platinum and gold for protons and alpha particles, which are shown in figure \ref{fig:Sp-Pt-Au} by solid lines (red and blue for the stopping power derived from the ELF obtained from Refs. \cite{Palik1999} and \cite{Werner2009}, respectively). 
Available experimental data are represented by letters \cite{PaulDatabase}. The symbols in the inset focus on the values around the stopping maximum and correspond to the most recent experimental data.
For comparison purposes, for the case of protons we have included the values of the stopping power from ICRU49 \cite{ICRU49} and from the semiempirical SRIM code \cite{SRIM2013}. For the case of alpha particles we have also added two more calculated stopping power curves \cite{Schiwietz2012,Montanari2013book}. As can be seen in the main figures, our calculations show small discrepancies with the experimental data at low projectiles energy $T$, surely due to the linear character of the dielectric formalism employed in our calculations. But in general, there is a rather good agreement with most of the experimental data in the whole energy range covered in our study. The behaviour around the stopping maximum is depicted in the insets of figure \ref{fig:Sp-Pt-Au}, where it can be seen that the agreement of our calculated stopping powers with the most recent experimental data \cite{Semrad1990,Eppacher1992,MartinezTamayo1996,Trzaska2002,Hsu2004,Zhang-Weber2005,Primetzhofer2012,Kumar2018,Trzaska2018,Moro2020,Selau2020} is particularly good. 

It is worth to notice that the results obtained from both sets of experimental ELF \cite{Palik1999,Werner2009} are practically identical. This is due to the fact that the stopping power is a quantity integrated over the momentum $\hbar k$ and energy $\hbar \omega$ transfers, which smooths the possible differences, as far as the ELF is well constructed (i.e., with physically motivated arguments).

\section{Summary and conclusions}
We have used the MELF-GOS methodology \cite{HerediaAvalos2005PRA,GarciaMolina2011} to obtain the energy loss functions of the transition metals Pt and Au from two different sets of experimental electronic excitation spectra \cite{Palik1999, Werner2009}.
The ELF obtained from these two sets of experimental data are not identical, but reproduce the main features of the excitation spectra of platinum and gold and fulfil physically motivated sum rules. 

The stopping power of platinum and gold for protons and alpha particles have been derived from these ELF by means of the dielectric formalism, taking into account the charge state fractions of the projectiles, the contribution from electronic capture and loss, as well as the polarisation of the projectile's electronic cloud due to the electric filed induced in the target. Our results are in rather good agreement with the available experimental data, the comparison being excellent with the most recent ones around the maximum stopping power. Besides, we have calculated the mean excitation energies for these metals, which are closer to the most recent data provided in a recent ICRU Report \cite{ICRU-Report73} than to other reported values.

\ack
We thank financial support from the European Union’s Horizon 2020 Research and Innovation programme under the Marie Sk{\l}odowska-Curie grant agreement no. 840752, the Spanish Ministerio de Econom\'{i}a y Competitividad and the European Regional Development Fund (Project no. PGC2018-096788-B-I00), and the Fundación S\'eneca (Project no. 19907/GERM/15).

\section*{References}
\bibliography{library}


\end{document}